\documentclass{appolb}

\usepackage{graphicx}
\usepackage{amssymb}
\usepackage{hyperref}
\usepackage[utf8]{inputenc}

\newcommand{\br}[1]{\langle #1\rangle}

\begin{document}

\title{Rapidity fluctuations in the initial state\thanks{e-mail: Wojciech.Broniowski@ifj.edu.pl, Piotr.Bozek@fis.agh.edu.pl}
\thanks{Research supported by the National
Science Centre grants DEC-2012/05/B/ST2/ 02528 and DEC-2012/06/A/ST2/00390.}}

\author{Wojciech Broniowski$^{1,2}$ and Piotr Bo\.zek$^3$
\address{$^{1}$Institute of Physics, Jan Kochanowski University, PL-25406~Kielce, Poland}
\address{$^{2}$The H. Niewodnicza\'nski Institute of Nuclear Physics, Polish Academy of Sciences, PL-31342~Cracow, Poland}
\address{$^{3}$AGH University of Science and Technology, Faculty of Physics and Applied Computer Science, PL-30059~Cracow, Poland}}

\maketitle

\begin{abstract}
We analyze two-particle pseudorapidity correlations in a simple model, where strings of fluctuating length are attached to wounded nucleons.
The obtained straightforward formulas allow us to understand the anatomy of the correlations, i.e.,  to identify the component due to the fluctuation of the number
of wounded nucleons and the contribution from the string length fluctuations.
Our results reproduce qualitatively and semiquantitatively the basic features of the recent correlation measurements at the LHC. 
\end{abstract}


PACS: 5.75.-q, 25.75Gz, 25.75.Ld

\bigskip

\begin{figure}
\vspace{-7mm}
\begin{center}
\includegraphics[width=0.5 \textwidth,angle=90]{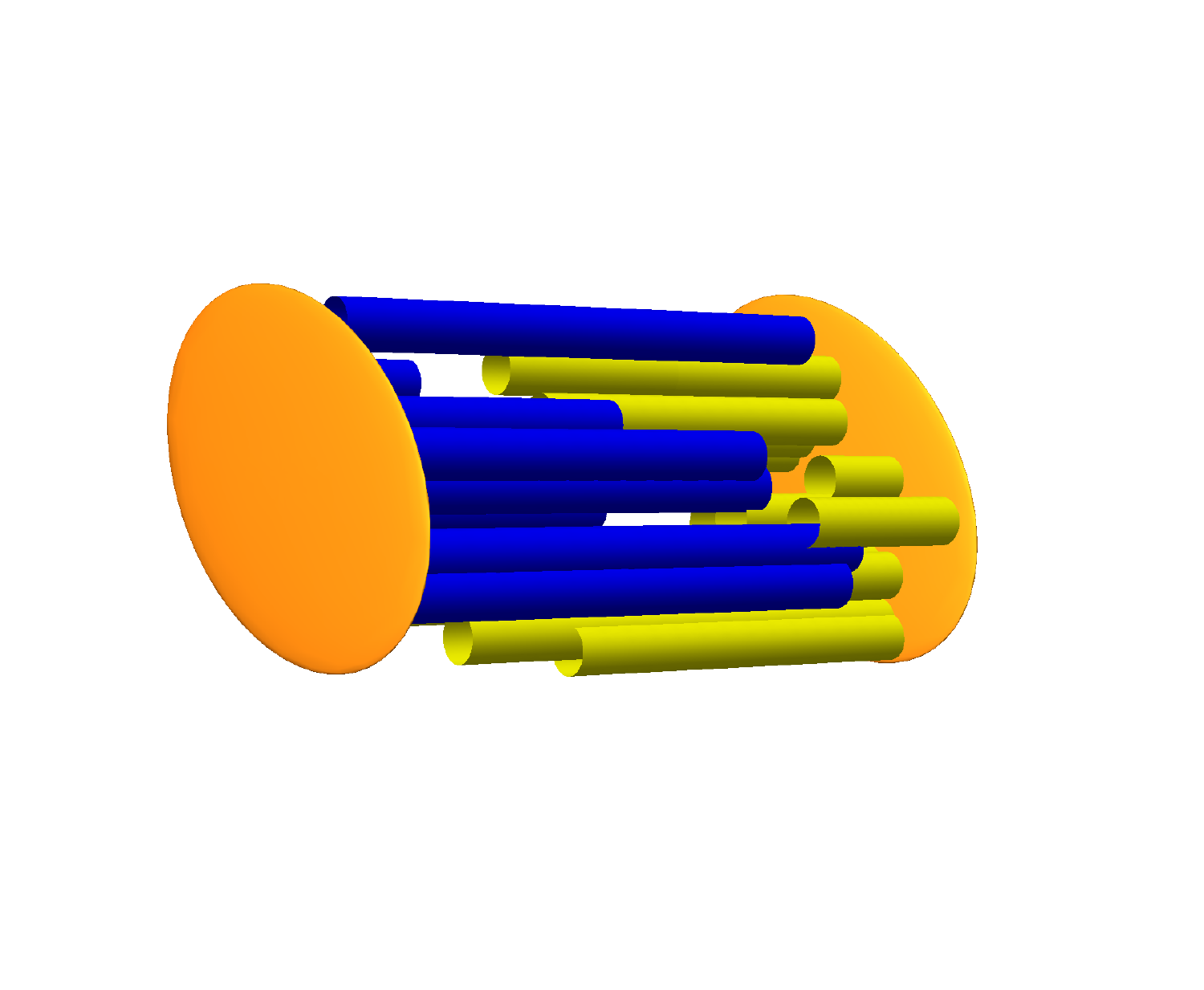} 
\end{center}
\vspace{-22mm}
\caption{Strings of fluctuating end-points extending  along the space-time rapidity. \label{fig:tubes}}
\end{figure}

In this talk (for details see Ref.~\cite{Broniowski:2015oif}) we analyze the recently measured two-particle pseudorapidity 
correlations~\cite{ATLAS:2015kla,ATLAS:anm} in a simple model with wounded nucleons pulling strings of fluctuating length.
The charges pulling the strings can be interpreted as the wounded nucleons~\cite{Bialas:1976ed}  (cf. Fig.~\ref{fig:tubes}). Importantly, the longitudinal position of the other end-point of the string 
is randomly distributed over the available space-time rapidity range. 
The assumption that the initial entropy distribution is generated by strings whose end-points are randomly
distributed is similar to the mechanism of Ref.~\cite{Brodsky:1977de}. It was also used in modeling 
the particle production in the fragmentation region~\cite{Bialas:2004kt}.

The two-particle correlation is defined as 
\begin{eqnarray}
C(\eta_1, \eta_2) &=& \frac{ \br{\rho(\eta_1,\eta_2)} }{ \br{\rho(\eta_1)}  \br{\rho(\eta_2)}}, \label{eq:C}
\end{eqnarray}
where $\rho(\eta_1,\eta_2)$ and $\rho(\eta)$ denote the distribution functions for pairs and single particles and $\br{.}$ stands for averaging over events. 
To minimize spurious effects, the ATLAS Collaboration~\cite{ATLAS:2015kla} uses a measure obtained from $C(\eta_1, \eta_2)$ by dividing it with its marginal projections, namely
\begin{eqnarray}
C_N(\eta_1, \eta_2) = \frac{C(\eta_1, \eta_2) }{C_p(\eta_1)C_p(\eta_2)}, \label{eq:CN}
\end{eqnarray}
where $C_p(\eta_{1,2})=\int_{-Y}^Y  d\eta_{2,1} \, C(\eta_1, \eta_2)$ and
$Y=2.4$ determines the experimental acceptance range for  $\eta_{1,2}$. 
Moreover, the correlation functions are conventionally normalized to unity, 
\begin{eqnarray}
 \overline{C}(\eta_1, \eta_2)  =  \frac{C(\eta_1, \eta_2)}{\int_{-Y}^Y \! d\eta_1 \int_{-Y}^Y \! d\eta_2 \,C(\eta_1, \eta_2) }, \label{eq:Cbar}
\end{eqnarray}
and analogously for $\overline{C}_N(\eta_1,\eta_2)$.

\begin{figure}
\begin{center}
{\includegraphics[width=0.65\textwidth]{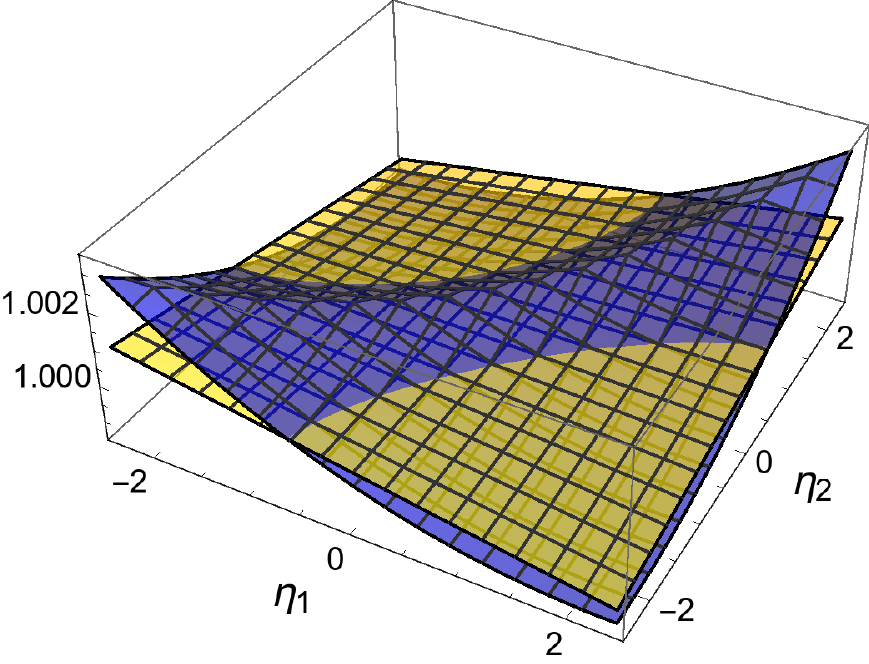}}
\end{center}
\vspace{-5mm}
\caption{Correlation function $\overline{C}_N(\eta_1,\eta_2)$ for Pb-Pb collisions at 2.76~TeV for $c=30-40\%$. 
The flat (lighter color) sheet corresponds to the calculation without the length fluctuations. The sheet with an elongated ridge (darker color) 
corresponds to the case with the string length fluctuations. \label{fig:c3040}}
\end{figure}

The details of the derivation of the formulas for the correlation function listed below are presented in Ref.~\cite{Broniowski:2015oif}. 
The number of wounded nucleons in the two colliding nuclei is denoted as $N_A$ and $N_B$, and $N_\pm=N_A\pm N_B$.
The cases with $r=1$ and $r=0$ correspond to, respectively, present or absent string length fluctuations. We introduce the scaled rapidities
\begin{equation}
u_{1,2}=\eta_{1,2}/y_b, \label{eq:u}
\end{equation}
with  $y_b$ denoting the rapidity of the beam. Then in the general case
\begin{eqnarray}
&&  \hspace{-5mm} C(\eta_1, \eta_2) = 1 +  \frac{1}{\left[ \br{N_+} + \br{N_-}u_1 \right] \left[ \br{N_+}+ \br{N_-}u_2 \right]} \label{eq:gen2}  \times \\ 
&&  \hspace{-2mm} \Big\{ \br{N_+} \left [ r (1 - {u_1 u_2} - |u_1-u_2|) +s(\omega)(1+r + (1-r)u_1 u_2  - r |u_1-u_2| ) \right ]  \nonumber \\ 
&&  \hspace{-3mm} + \br{N_-} s(\omega) (u_1+u_2)  + {\rm var}(N_+)  + {\rm var}(N_-) u_1 u_2  + {\rm cov}(N_+,N_-)(u_1+u_2) \Big\}, \nonumber
\end{eqnarray}
whereas for symmetric collisions ($A=B$)  the formula simplifies into 
\begin{eqnarray}
&& \hspace{-5mm} C(\eta_1, \eta_2) = 1 +  \frac{1}{\br{N_+}^2} \label{eq:gensym2} \times \\
&& \Big\{ \br{N_+} \left [ r( 1 - {u_1 u_2} -  |u_1-u_2|) +s(\omega)(1+r + (1-r)u_1 u_2  - r |u_1-u_2| ) \right ]  \nonumber \\
&& + {\rm var}(N_+)  + {\rm var}(N_-) u_1 u_2 \Big\}. \nonumber
\end{eqnarray}
The quantity $s(\omega)$ stands for the square of the standard deviation of the overlaid distribution of strength of the sources.

\begin{figure}
\begin{center}
{\includegraphics[width=0.65\textwidth]{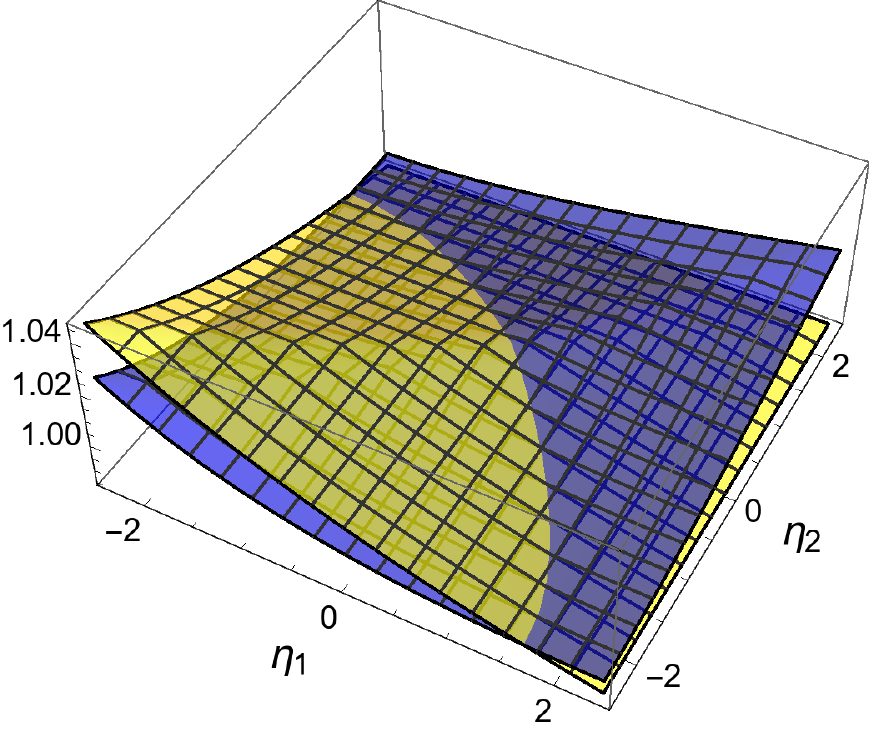}}
\end{center}
\vspace{-5mm}
\caption{Correlation functions $\overline{C}(\eta_1,\eta_2)$ (lighter color)  and $\overline{C}_N(\eta_1,\eta_2)$ (darker color) for 
p-Pb collisions at 5.02~TeV. \label{fig:pPb}}
\end{figure}

As originally noticed in Ref.~\cite{Bzdak:2012tp}, fluctuation in the number of wounded nucleons alone ($r=0$, $s(\omega)=0$) generates non-trivial 
longitudinal correlations. Our formula shows, however, that a significant (and long-range in rapidity) part comes from the length fluctuations.
The results are depicted in Fig.~\ref{fig:c3040}. Results of a similar study for the asymmetric case of p-Pb collisions are shown in Fig.~\ref{fig:pPb}.

Analytic expressions may be obtained \cite{Broniowski:2015oif} for the $a_{nm}$ coefficients of the expansion in a set of orthonormal polynomials~\cite{Bzdak:2012tp,Jia:2015jga}

The features found in our simple model are manifest in advanced 
models implementing string decays in the early phase of the high-energy collisions~\cite{Andersson:1983ia,Wang:1991hta,Lin:2004en,Monnai:2015sca}.
Our formulas provides an intuitive understanding for these mechanisms. 

\bigskip

We cordially wish Janek Pluta all the best on the occasion of his anniversary. Let the successful Cracow-Warsaw collaboration, animated by Janek long ago, continue 
for many years to come.

\bibliography{hydr}

\end{document}